\documentclass{appolb}
\usepackage{graphicx}

\begin{document}
\title{Specific Heat Study on Heavy-Fermion Pr compounds\\
 with filled skutterudite structure%
\thanks{Presented at the Strongly Correlated Electron Systems 
Conference, Krak\'ow 2002}%
}


\author{Y. Aoki, T. Namiki, T. Kanayama, S. Ohsaki, S. R. Saha, H. Sugawara, H. Sato
\address{Dept. of Physics, Tokyo Metropolitan University, Tokyo 192-0397, Japan}
}


\maketitle


\begin{abstract}
Rare $4f^2$-based heavy-fermion behaviors have been revealed recently in Pr-based filled skutterudites PrFe$_4$P$_{12}$ and PrOs$_4$Sb$_{12}$.
Recent studies on the thermal properties on both compounds are reported, putting emphasis on the field-induced ordered phase found in PrOs$_4$Sb$_{12}$.
\end{abstract}



\section{Introduction}
The 4$f^2$ configuration of Pr ions in intermetallic compounds had been considered to be quite stable in view of no observation of strongly correlated electron behaviors until recent discovery of heavy fermion (HF) behaviors in some of Pr-based compounds.
Such unprecedented examples are PrInAg$_2$~\cite{Ya96} and filled skutterudites PrFe$_4$P$_{12}$~\cite{AoPrFe4P12PRB} and PrOs$_4$Sb$_{12}$~\cite{BauerPRB}.
In the latter group, not all Pr-based compounds show such unconventional behavior as summarized in Table~\ref{table:1} and at this stage it is not clear whether there is a systematic trend or not.
In this article, we report recent studies on the thermal properties in the two Pr-based compounds and discuss the nature of the HF ground states.
\begin{table}
\caption{Thermal properties of PrT$_4$X$_{12}$. See text for abbreviation. Roughly estimated $\gamma$ (or $\Delta C/T$ at phase transitions) are given in the unit of J/K$^2$mol.} 
\begin{tabular}{c|ccc}
\hline
\multicolumn{1}{c|}{\textbf{T \textbackslash X}} & \multicolumn{1}{c}{P} & \multicolumn{1}{c}{As} & \multicolumn{1}{c}{Sb} \\
\hline
\hline
Fe & 
\begin{tabular}{@{}p{10em}@{}} HF $\gamma \sim 1$ in 4 T \\ AFQ? $T_Q$=6.5 K \cite{AoPrFe4P12PRB} \end{tabular} &
\begin{tabular}{@{}p{7em}@{}} ------------------ \end{tabular} &
\begin{tabular}{@{}p{9em}@{}} Ferromagnet? \\ $T_m$=5 K \cite{DaneJPCS,BauerPB} \\ $\gamma \sim 1$ in 3 T \end{tabular} \\ 
\hline
Ru & 
\begin{tabular}{@{}p{10em}@{}} M-I transition \\ $T_{MI}$=62~K~CDW? \\ $\Delta C/T_{MI}=0.2$ \cite{SekinePB,MatsuhiraPB} \end{tabular} & 
\begin{tabular}{@{}p{7em}@{}} SC $T_c$=2.4 K \\ $\gamma=0.1$ \cite{ShiroPRB,NamikiIP} \end{tabular} &
\begin{tabular}{@{}p{9em}@{}} SC $T_c$=1.08 K \\ $\gamma=0.06$ \cite{TakedaJPSJ} \end{tabular} \\
\hline
Os & 
\begin{tabular}{@{}p{10em}@{}} no ordering \\ above 2 K \cite{SekinePRL} \end{tabular} &
\begin{tabular}{@{}p{7em}@{}} ------------------ \end{tabular} &
\begin{tabular}{@{}p{9em}@{}} HFSC T$_c$=1.8 K \\ $\Delta C/T_c=0.5$ \cite{BauerPRB,AoPOS} \end{tabular} \\
\hline
\end{tabular}
\label{table:1}
\end{table}

\section{HF behavior and anomalous ordered state in PrFe$_4$P$_{12}$}

In PrFe$_4$P$_{12}$, HF behavior is evidenced by the largely-enhanced electronic specific-heat-coefficient ($\gamma \sim 1$ J/K$^2$mol)~\cite{AoPrFe4P12PRB} and cyclotron masses in dHvA studies~\cite{SuPrFe4P12} in applied fields where an anomalous ordered state (ODS), appearing below $T_Q=6.5 $ K, is suppressed.
Non-magnetic nature of the ODS has been revealed by the $^{141}$Pr-nuclear specific heat and nuclear scattering studies~\cite{Keller,AoSCES2001}.
A possible order parameter of the ODS is an antiferro-quadrupole (AFQ)~\cite{CurnoeSCES2001}, which couples to the observed lattice distortion~\cite{IwasaSCES2001} and consequently to the expected Fermi surface instability~\cite{HarimaSCES2001}.
In this scenario, a gap opening suggested from the electrical resistivity $\rho (T)$ and Hall coefficient $R_H (T)$~\cite{SaPrFe4P12PRB} and a field-induced staggered magnetic component observed in neutron scattering experiments~\cite{Iwasa2002} can be naturally understood.
This scenario might point to a possibility that the quadrupole fluctuation is essential for the HF behavior in the high fields.

We have extended our study to La-substituted system (Pr$_x$La$_{1-x}$)Fe$_4$P$_{12}$.
It is found that the ODS smears out rapidly with the La doping and a new magnetic transition appears for $x \le 0.95$ at $T_m=1 \sim 2$ K (depending on $x$), suggesting an competition between the quadrupole and the magnetic interactions in this compound (reported elsewhere).
Interestingly, after the suppression of the ODS by the La doping, the HF behavior is recovered even in zero field and the $\gamma$ value below $T_m$ is $\sim 1.7 $ J/K$^2$mol-Pr at $x=0.85$~\cite{aokiJPCS}, indicating the compatibility between the magnetic ordering and the HF behavior.

\section{Field-induced ordered state in HFSC PrOs$_4$Sb$_{12}$}

Compelling evidence for the HF superconductivity (HFSC) in PrOs$_4$Sb$_{12}$ was given by a large specific heat jump $\Delta C/T=0.5$ J/K$^2$ mol at $T_c=1.85$ K on a pellet of compressed powdered single crystals~\cite{BauerPRB}. 
The jump is superimposed on a Schottky-like anomaly appearing at $\sim 3 $ K, which could be interpreted as a crystalline-electric-field (CEF) excitation or a strongly energy-dependent quasiparticle excitation.

In order to study the nature of the anomalous normal state, we have measured specific heat in magnetic fields using a single crystal of high quality, which has been ensured by an observation of the dHvA oscillations~\cite{SugawaraSCES2002}.
The Schottky-like anomaly was found to be drastically suppressed with increasing magnetic field, i.e., the downward curvature in $C(T)$ disappears in 5 T or higher \cite{AoPOS}, indicating that the low-lying excitation leading to the anomaly has a magnetic character.
\begin{figure}[!ht]
\begin{center}
\includegraphics[width=0.7\textwidth]{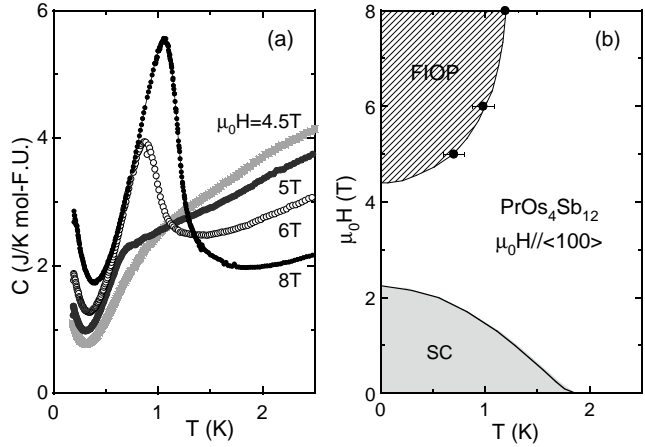}
\end{center}
\caption{(a) Temperature dependences of specific heat $C(T)$ in different fields and (b) magnetic-field-vs-temperature phase diagram of PrOs$_4$Sb$_{12}$. Partly from Ref.~\cite{AoPOS}.}
\label{fig}
\end{figure}

Another important finding in our $C(T)$ data is a clear thermodynamical evidence for the existence of a field-induced ordered phase (FIOP) above 4.5 T.
As shown in Fig.~\ref{fig} (a), a distinct anomaly evidencing a field-induced phase transition develops with increasing field.
The transition temperature $T_x$ is plotted in a $\mu_0 H$-vs-$T$ phase diagram of Fig.~\ref{fig} (b).
Below 8 T, $dT_x/dH$ is positive and the Ehrenfest's theorem suggests that magnetization $M$ is enhanced in the FIOP.
This behavior is actually reflected in the field dependence of the $^{141}$Pr-nuclear contribution \cite{AoPOS}, which causes the upturn in $C(T)$ below 0.5 K (see Fig.~\ref{fig} (a)).
We speculate that the order parameter is of an AFQ accompanied by a field-induced antiferromagnetic (AF) component, as observed in CeB$_6$~\cite{CeB6Sera2001} and TmTe~\cite{TmTeLink1998}, since AF correlations inferred from $\chi(T)$ could energetically stabilize the FIOP in magnetic fields leading to $dT_x/dH>0$.

Bauer {\it et al.} pointed out a possibility that the HF behavior in PrOs$_4$Sb$_{12}$ involves a quadrupole Kondo interaction on Pr ions based on their CEF model, which has a non-Kramers doublet ground state~\cite{BauerPRB}.
From this view point, it is essentially important to clarify whether any residual entropy is hidden below 0.1 K since, for example, a residual entropy of $1/2 R \ln 2$ is predicted for a single-site quadrupole Kondo model~\cite{CoxZawad98}.
To check this, we performed quantitative measurements of magnetocaloric effect, which have worked satisfactorily to study the thermal properties of the metamagnetic-like anomaly in CeRu$_2$Si$_2$~\cite{AoCeRu2Si2}.
No anomalous entropy ($ > 3 \times 10^{-3} R \ln 2$) has been detected in $\mu_0 H<8 $T \cite{AoPOS}.
We cannot deny, however, a possibility that Pr ions in PrOs$_4$Sb$_{12}$ form a quadrupole Kondo lattice with an effective singlet ground state.

Determination of the order parameter of the FIOP is essential to understand the anomalous nature of the normal state, which provides the background of the rare $4f^2$-based HFSC in PrOs$_4$Sb$_{12}$.

This work is supported partly by a Grant-in-Aid for Scientific Research from the Ministry of Education, Science and Culture.

\end{document}